\documentclass[12pt,epsf]{article}

\usepackage{amssymb,amsmath}

\newcommand{\ct}{\cite}

\newcommand{\bc}{\begin{center}}
\newcommand{\ec}{\end{center}}
\newcommand{\bd}{\begin{displaymath}}
\newcommand{\ed}{\end{displaymath}}
\newcommand{\be}{\begin{equation}}
\newcommand{\ee}{\end{equation}}
\newcommand{\ba}{\begin{array}}
\newcommand{\ea}{\end{array}}
\newcommand{\bea}{\begin{eqnarray}}
\newcommand{\eea}{\end{eqnarray}}
\newcommand{\bt}{\begin{tabular}}
\newcommand{\et}{\end{tabular}}

\newcommand{\bp}{\begin{picture}}
\newcommand{\ep}{\end{picture}}
\newcommand{\bfi}{\begin{figure}}
\newcommand{\efi}{\end{figure}}

\begin{document}

\title{\huge {\bf  Dark Matter Balls Help
Supernovae to Explode }}

\author{ C. D. Froggatt ${}^{1}$ \footnote{\large\, colin.froggatt@glasgow.ac.uk} \ \ \
H.B.~Nielsen ${}^{2}$
\footnote{\large\, hbech@nbi.dk} \\[5mm]
\itshape{${}^{1}$ Glasgow University, Glasgow, Scotland}\\[0mm]
\itshape{${}^{2}$ The Niels Bohr
Institute, Copenhagen, Denmark}}

\maketitle

\begin{abstract}
As a solution to the well-known problem
that the shock wave potentially responsible
for the explosion of a supernova actually tends to
stall, we propose a new energy source arising
from our model for dark matter.
Our earlier  model proposed that dark
matter should consist of cm-large white
dwarf-like objects kept together by a
skin separating two different sorts
of vacua.
These dark matter balls or pearls will
collect in the middle of any star throughout its lifetime.
At some stage during the development of a supernova
the balls will begin to take in neutrons and then
other surrounding material.
By passing into a ball nucleons fall through a potential
of order 10 MeV, causing a severe production of heat - of
order 10 foe for a solar mass of material eaten by the balls.
The temperature in the iron core will thereby be raised, splitting
up the iron into smaller nuclei. This provides a mechanism for
reviving the shock wave when it arrives and
making the supernova explosion really occur.
The onset of the heating due to the dark
matter balls would at first stop the collapse of
the supernova progenitor.
This opens up the possibility of
there being {\em two}
collapses giving two neutrino outbursts,
as apparently seen in the supernova
SN1987A - one in Mont Blanc, and one 4 hours 43
minutes  later in both IMB and Kamiokande.
 \end{abstract}\date{}


\newpage

\thispagestyle{empty}
\section{Introduction}
A supernova explosion is supposed to
originate from in-falling material of the
progenitor-star
being reflected after having been stopped
by the nuclear forces, when  a neutron
star is first formed and compressed to about
double nuclear matter density \ct{Rolf,Bethe,Woosley}. The
re-expansion of the compressed
neutron star in the center would then cause
a shock wave to propagate outward. This
shock wave is expected to cause
what is seen as the supernova explosion.
However, more detailed calculations
suggest that, at least unless one includes
convective or non-symmetric development,
the shock wave tends to stall before
reaching out far enough to expel the stellar
envelope and provide sufficient energy for
the observed magnitude of supernova
explosions.

This conclusion that an insufficient amount
of energy is deposited into the material expelled
from the core remains true, even
when the effect of a flux of neutrinos
from the center is included in the calculations.
Heating from these neutrinos does though not revive
the shock wave sufficiently to provide the
energy of $1 \ \hbox{foe} \equiv 10^{51} \
\hbox{ergs}$ needed by the observed stellar remnants and
radiation. It is not that there is
insufficient energy available in the
collapse, because the gravitational collapse to the
neutron star easily releases 100 foe.
Nevertheless the simulations show that
the shock wave emitted
runs out of force and cannot even
provide the one foe needed \ct{Papish}.

It is still hoped that more detailed two
dimensional or three dimensional
simulations including convection
could explain how, at least in some
direction, enough
energy would be brought to revive
the shock wave so as to provide the
observed explosion \ct{Janka}. Alternatively
some extra source of energy providing
this ``revival'' could help \ct{foot}.

It is indeed such an {\em extra} energy
source, which we propose in the present
article.
In section
\ref{sec2}
below we shall describe our special model
\ct{crypto1,crypto2,Tunguska} for dark
matter, which has the peculiar property
that it can unite with ordinary matter and thereby release an energy of the order
of 10 MeV per nucleon. In fact one should
think of our dark matter as consisting of
pearls of cm-size with an interior in
which there is a different type of vacuum.
When nucleons penetrate into this
new type of vacuum, it is supposed that
they pass a potential barrier so as to
release for heat production about 10 MeV
per nucleon.
It should, however, be noticed that these
pearls are supposed to be surrounded by
a thin region in which there is an
electric field preventing say protons
and, even more so, heavier nuclei from
penetrating into the pearls. So {\em
only} when the protons or the nuclei
have got sufficiently high temperature
to pass this electrical field will
the pearls begin to take them up from
the surrounding material. Neutrons,
however, may be able to penetrate even
at low temperature.

In the following section \ref{sec2}
we shall review our model for dark
matter as being some very heavy pearls,
with a mass of about $10^8 \ \hbox{kg}$ per cm-size pearl.
In section \ref{sec3} we then give
the scenario for the development of
a core collapsing supernova, with special
emphasis on the activity of our dark
matter pearls.

In subsection \ref{sec4} we provide an
estimate of the time between the first
collapse of the iron core, which is interrupted
by the ignition of our pearls, and the second
final collapse. The importance of this
time difference is that apparently
{\em two} neutrino outbursts were
observed in the supernova SN1987A,
with a time difference
of 4 to 5 hours.

Finally in section \ref{sec5} we conclude
and resume.

\section{Our Dark Matter Model}
\label{sec2}
Usually it is believed that dark matter
must result from physics beyond the
Standard Model, e.g.~WIMPs \ct{pdg}
usually identified as the lightest
SUSY partner of the known particles.
Although ATLAS and CMS have looked for SUSY partners,
they only found lower limits for their masses \ct{Susy}.
Also claims for direct detection of WIMPs
\ct{Dama, Cogent, Cresst} are in contradiction
with experiments not having seen any \ct{Cdms1,Cdms2,Cdms3}.
We have, however, for some time been
working on a model \ct{crypto1,crypto2,Tunguska}
for the dark matter,
which does {\em not need an extension
of the Standard Model with new fundamental
particles}. Rather we propose in our
model, as new physics,  {\em only} some
bound states
composed of 6 top and 6 anti-top quarks
bound together mainly by Higgs-forces already
present in the Standard Model itself.
It should though be admitted that we
supplement the Standard Model with a
fine-tuning principle, the ``Multiple
Point Principle'' \ct{MPP1,MPP2,MPP3}, which is defined
to mean that the coupling constants in
the Standard Model are adjusted so as
to arrange for several - in fact we
think 3 - different vacua to have
just the same energy densities.
The existence of one of these speculated
vacua led us to the {\em pre}diction
of the Higgs mass \ct{tophiggs}. The existence of
another such speculated new vacuum
is supposed to lead
to the adjustment of mainly the top-Yukawa
coupling, so as to make 6 top plus
6 anti-top quarks bind very strongly
together and form a relatively
very light bound state \ct{nbs1,nbs2,nbs3,boundstate}.
Furthermore, according to our ``Multiple
Point Principle'', a vacuum is supposed to form
containing a condensate of this bound state
and having the same energy density as the vacuum
in which we live.

The idea of our model for dark matter
is now that the dark matter floating
around in space consists of small
pearls of cm-size, inside which is a
bubble of the bound state condensate
vacuum. Between the two different types
of vacua there will be a skin - a surface
tension one could say - of a rather high
density (because its order of magnitude
is give by the scale of weak interactions)
 $S \sim 4 *10^8 \ \hbox{kg/m$^2$}$. In order that such a
bubble be stabilized it has to be pumped
up with some material under sufficiently
high pressure to resist the pressure
from this skin. In our pearls making
up the dark matter this
pressure is about $5 * 10^{27} \ \hbox{N/m$^2$}$,
and the interior of the pearl is much
like a little white dwarf star with a
density of ordinary matter inside it
of the order of $10^{14} \ \hbox{kg/m$^3$}$
= $ 10^{11} \ \hbox{g/cm$^3$}$. A typical ball has a
radius of 0.67 cm and mass of order
$10^8$ kg. We note that our dark matter balls are too light for
observation by microlensing \ct{Machos1,Machos2,Machos3}.
In order to keep the ordinary
matter inside the pearls from expanding
out, it is crucial that
a nucleon feels a potential difference
in passing through the skin such that its
potential inside the pearls is lower by
about 10 MeV than outside. In fact we
estimated the potential difference to be
$\Delta V = 10 \pm 7$ MeV.

It is this potential difference of 10 MeV
that is crucial for our idea of using our
pearls to help the supernovae to truly
explode. This potential difference means that
an energy of 10 MeV per nucleon is released,
whenever a nucleon is brought inside
one of our pearls. Now, however, this
transport of nucleons into the interior
of the pearls is prevented, because these
pearls are normally surrounded by an
electric field repelling protons
as well as nuclei. This field is
there due to the fact that,
analogously to white dwarfs, the pearls
contain degenerate electrons. We expect
our typical pearls to contain a degenerate
Fermi sea of electrons with a Fermi energy
of the order of 10 MeV. While now the
protons are kept inside the skin by the
potential difference mentioned above,
the Fermi sea of electrons will spill
over to the outside of the skin region.
Thus, in a little range of order
$ 20 \  \hbox{fm} $ outside the skin,
there are electrons - but no
protons. This gives rise to
an electric field in much the same way
as there is an electric field inside an
atom, due to the protons in the nucleus
being charge-wise compensated only by
the electrons, which are placed
appreciably further out. It is the
electric field around the pearls, which prevents
protons and/or nuclei from penetrating
into the pearls. They have first to tunnel
or otherwise pass through this electric
field, before they can be caught by the
nuclear potential of the 10 MeV which we have
hypothesized. If the pearl gets bigger than
our typical radius of $0.7 $ cm the electric
field layer gets thicker, with
a thickness proportional to
the fourth root of the pearl radius;
but at the same time the electric field strength
becomes smaller the bigger the radius
and the potential for passing the electric
field layer varies as the inverse of the
fourth root of the radius. Thus it gets easier
for a positively charged particle (proton or nucleus)
to penetrate into the
ball, as the ball grows in size.

We take the size of typical pearls
to be close to the
critical point for collapse under their
pressure against the assumed 10 MeV
potential difference across the skin.
Then, taking the
dark matter density in our galaxy to be
$0.3 \ \hbox{GeV/cm$^3$}$, we estimate that the
earth is hit by one of our pearls
about once every 200 years, matching with
the assumption that the
famous Tunguska event \ct{Tunguskaevent} was caused by the
fall of such a pearl \ct{Tunguska}.
An impact rate of one pearl per 200 years
means that the earth should have been hit
by $2 * 10^7$ pearls in its history.
Correspondingly then the sun should have
been hit by
$2 * 10^7*(R_{sun}/R_{earth})^2 = 2*10^{11}$
pearls. Since each pearl
has a mass of $10^8$ kg, this means that the
sun should have
collected $2*10^{19}$ kg of dark matter, which is
$10^{-11}$ times the mass
of the sun. The supernova SN1987A was
supposedly about 20 times
heavier than the sun \ct{Arnett}, when it exploded.
Taking
the variation of the radius and lifetime
of a main sequence
star to vary respectively like
$R_{MS}\propto M^{0.78}$ and
$\tau_{MS} \propto M^{-2}$,
we get
$R_{MS}^2\tau_{MS} \propto M^{-0.5}$. Thus
$0.5*10^{11}$ pearls should have collected
in the SN1987A, with a collected mass of
$0.5 * 10^{19}$ kg, which is
$10^{-13}$ times the mass of the supernova.

\section{Scenario of Supernova
Collapse}
\label{sec3}
During mainly the main sequence development of
the supernova-star the dark matter
pearls in our model fall into the star, where
they get stopped and then fall essentially to the
center of the star. One should have in mind that
our pearls have
densities of the order of $10^{14} \ \hbox{kg/m$^3$}$.
At the relatively low temperature of
$10^7$ K $\approx1 \ \hbox{keV}$  in the center
of the star, during its
main sequence development, neither protons,
other nuclei nor the pearls themselves can pass
through the
electric field\footnote{As in our previous article \ct{Tunguska} we assume that the typical
pearls making up the dark matter have such a size that they are just on the borderline
of stability, where their protons would escape and the pearls would collapse.
In this case, the proton would have to just pass the total
potential difference for getting in or out of the pearl. Now we already
assumed that there is a potential drop of 10 MeV on passing through the skin of
the pearl. Thus, in order to make the {\em total} potential difference for the proton
in passing from inside to outside the pearl zero, the electric
potential has also to be 10 MeV. For pearls bigger than
the ``critical size'' the electric potential
difference will be smaller.} surrounding a pearl with
a potential difference typically
of the order of
10 MeV. So at this time the pearls are
quite inert.

However, the pearls can begin to sip up
material and expand, if the charged
particles in the
surroundings get so energetic/hot
that they can penetrate
the electric field and then gain
energy from passing
into the pearls through the skin.
Alternatively free neutrons may become available
and they can just pass
through the electric field
without problems, because they are
neutral. In the actual supernova,
it is the absorption of neutrons
which becomes important first.
This causes a very rapid expansion of the
pearls after the silicon burning era
when the supernova progenitor begins its
Kelvin-Helmholtz gravitational collapse.
The consequent huge deposition of
energy of 10 MeV per absorbed nucleon
halts the collapse of the star until the interior
of the star cools down again. In this way
we obtain a two stage gravitational collapse \ct{deRujula},
with possibly an intense burst of neutrinos at
each stage.

\subsection{Absorption of Material by Balls}
\label{absorption}
The speed $v_{wall}$ with which the pearl skin
or wall comes to move relative to the
surrounding material is estimated in the
following way:

It is only the neutrons that pass freely
into the pearls.
Supposing their number density (outside the pearls) is $n_n$
and that their thermal
speed is $v_n$; then a layer of neutrons
of thickness $v_n$
penetrates into the pearl every time
unit. Supposing now
that the density of nucleons inside the
pearls is in
our model $10^{14} \ \hbox{kg/m$^3$}
= 10^{11} \ \hbox{g/cm$^3$}
\sim 6*10^{34} \ \hbox{nucleons cm$^{-3}$}$,
 then the speed
with which the pearl
expands, without having to
change its density, becomes
\begin{equation}
 v_{wall} =
v_n *\frac{n_n}{6*10^{34} \ \hbox{cm$^{-3}$}}
\label{vwall}
\end{equation}

For instance at the silicon burning time,
when silicon burns
into elements in the iron group, the
temperature is of the order of
$4 *10^9 \ K
= 0.4 \ \hbox{MeV}$
\cite{Rolf}. Thus
the speed of say a neutron is then
$\sqrt{3* 0.4\ \hbox{MeV}/\hbox{GeV}} c
= 5*10^{-2} c
= 10^7\ \hbox{m/s}$
At this time the density is
$3*10^7\ \hbox{g/cm$^3$}$ and,
  from Bodansky et al \cite{Fowler}, the
density
of neutrons is given as
$n_n= 10^{20}\ \hbox{cm$^{-3}$} = 10^{26}\ \hbox{m$^{-3}$}$.
Thus
the speed of the wall becomes
$v_{wall} = 10^7 *10^{26}/6*10^{40}\ \hbox{m/s}
= 1.7 *10^{-8}\ \hbox{m/s}$. With the time scale
taken as the one
day it takes to pass through the silicon burning era,
the skin would
have moved just $0.9*10^5*1.7 *10^{-8}\ \hbox{m}
= 1.5\ \hbox{mm}$.

Normally there is an electric field
in a thin layer around a dark matter pearl
pointing perpendicularly to the
skin or the surface separating the two vacua.
This is so because there is normally a difference in
density - for instance of electrons - between the two sides
of the surface. This electron density falls off
gradually across the surface of separation, while the
proton density typically changes very abruptly at the
skin. The latter has a very small thickness
compared to atomic physics scales. If the density variation
of the electrons and the protons do not follow each other closely
across the surface, an electric field will appear
at this surface.
If, however, the density of matter inside
and outside our pearls would be
the same, so that especially the density of
electrons would be the same
on the two sides, there would  be no electric field.

Thus, once the
density in the outside becomes of the order of the density
$10^{14}\ \hbox{kg/m$^3$}$ inside our pearls, the skin can move
rather freely; it would do so with
essentially the thermal speed
of the particles. So, when the density in
the surroundings becomes of this
order, the pearls would expand rapidly
even without the need
for any neutrons.
Assuming that the density of the material
at the center of the star is given by
the relationship \cite{Woosley}
$\rho = 10^6 T_9^3$, where $T_9$ is the temperature
in units of $10^9\ K$, our
pearl-density is achieved for the temperature
$T_9 = (10^{11}\ \hbox{g/cm$^3$} /10^6)^{1/3}
= 50 $.
Using instead the estimate of the temperature
of star matter of density
$10^{11}\ \hbox{g/cm$^3$}$
given in
\cite{Bethe,Cooperstein}, we get
$ T = 1.21\ \hbox{MeV} = 1.2 *10^{10}\ K$; this means
$T_9 = 12$, which is four times smaller than our first estimate
because it includes the effect of
some decrease in entropy towards the core.
So if we wait for even charged matter to be sipped up,
it would start in the temperature range $T_9 = 12 \ \hbox{to} \ 50$.

But the presence of free
neutrons in the surrounding
matter becomes sufficiently copious
at a lower temperature and
causes a rapid development
of the pearls, before
the density gets sufficiently high for the absorption of
charged particles.
In fact the free neutron density becomes a few percent
of the total star density for total densities of
$10^{10}\ \hbox{g/cm$^3$} $ and above \ct{Bethe,Cooperstein}.
It follows that, when the total star density is of
this order of $10^{10}\ \hbox{g/cm$^3$}$, neutrons are absorbed
by the pearls with a significant rate
for the supernova. In fact, from equation \ref{vwall}, the
speed $v_{wall}$ of the
skin of the pearls relative to the surrounding material
becomes of the order of $10^{-3}$ times
the thermal speed.
If the temperature now were say 1 MeV,
then the
thermal speed would be $\sim c/18$ and thus
the speed of the
wall would be $10^{-3}*3 *10^8\ \hbox{ms$^{-1}$} /18
 \sim 10^4\ \hbox{m/s}$,
meaning that a region with radius of the order of
100 km would be
passed in 10 s. So, under these
conditions, the pearl
would spread explosively in a few seconds.

This fast absorption might though be
damped by
the pearls picking up the neutrons and
thus becoming damped
in their expansion, until the previous neutron density
 in their neighborhood has
been essentially re-established by nuclear statistical equilibrium
in the star. The pearls cannot pick up
the protons as
long as there is an electric field present,
which is expected to
be there until the density is the same
on both sides
of the wall.

In spite of such possible damping
effects, we still
believe that the expansion of the
pearls can easily
become fast enough that we must
consider the
process explosive. Thus the whole region
in which the absorption goes on
gets strongly heated and essentially all
the energy from the
passage of the nucleons (as neutrons we
suppose), each delivering
10 MeV, gets collected in such a
region.

We expect  the expansion to stop, when the density
of the material surrounding the balls becomes sufficiently
low.

\subsection{Calculation of Time
Interval between the Neutrino
Emissions}
\label{sec4}
When
the explosive expansion
of the pearls takes place,
an energy of $10 \pm 7$ MeV per
nucleon (the uncertainty is estimated in \ct{Tunguska})
passing into the pearls is released. This will
lead to a
temperature increase in the region over
which this released energy
is getting distributed.
Now a thermal energy of 10 MeV obtained
by a nucleon will get distributed as
$\Delta T$ for each
degree of freedom. Thus at first it seems
that the temperature will be raised
by $ \Delta T = 10\ \hbox{MeV}/3 = 3\ \hbox{MeV}.$\footnote{This
estimate would be correct if the
potential in the plasma for the nuclei were treated
as a harmonic oscillator potential. If, however,
we counted the particles as free particles
only the kinetic energy would be non-zero and
we would instead have argued for
$\Delta T = 10\  \hbox{MeV} /(3/2)
= 6 \ \hbox{MeV}$ at this stage.
Presumably $\Delta T$ lies in between 3MeV and 6 MeV.}
We are interested in the mean
excess temperature in the period until
the released energy
has been emitted out of the region by essentially
neutrinos. The mean excess temperature
can of course be at most
half of the starting value.
This means it will at most be
$\Delta T = 10\ \hbox{MeV}/(3 *2) = 1.7\ \hbox{MeV}$.

Let us imagine the situation just after
the explosive expansion of the balls took place:
\begin{enumerate}
\item  The pearls/balls have most likely
united together in one big ball, with its center approximately
coinciding with the center of the star.
\item Of course the interior of this
united ball has been heated up by the extra
temperature $\Delta T$. But, in addition,
there is a region outside - and thus above
the ball - in which a similar temperature
increase has been caused by the dark matter
explosion.  We may guess that this outside
region, which is similarly heated, is
of about the same mass as the united ball.
\item The explosion itself may have only taken
say 10 seconds.
\item But now the cooling, by neutrinos mainly \ct{Munakata,Itoh},
sets in. In first approximation this cooling rate is just
given, as we shall see below, to be of the order
of   $10^{14}\ \hbox{erg/g/s}$. However, it is
likely that the cooling is a bit slower
further away from the center than deeper down, because the density
higher up in the core is somewhat lower.
Also presumably most of the heat will be produced around the skin of the ball.
Both these effects could easily lead to the central region reaching
the temperature and pressure, where the Kelvin-Helmholtz gravitational
collapse restarts, first.
\item The Kelvin-Helmholtz collapse restarts in the central region
when the extra temperature $\Delta T$ has been cooled away by neutrino
emission; we estimate below this cooling should take of the order of 14 hours.
However the upper part of the region heated by pearl expansion and
nucleon absorption will still remain somewhat heated up compared to
the central region.
\item For instance, in this higher region, the iron peak nuclei could
still be split into say nucleons or at least helium.
 \end{enumerate}

In order to estimate the cooling time for the
excess temperature $\Delta T$ to be dissipated,
we should estimate the ratio $r $ of the total amount of matter
significantly heated to the amount of matter sucked in
by the pearls and finally contained in the big ball.
Under point 2 in the list above we guessed that the amounts of
matter heated inside and outside the ball were about the same.
Taking the initial pearls to make up a negligible amount of mass,
this means that we take the ratio $r$ to be 2.
Thus twice as many nucleons as at first thought are heated up
and the temperature increase gets reduced by a factor of 2 compared
to our first estimate above of $\Delta T =1.7$ MeV.
This means we get the true temperature increase to be
$\Delta T = 1.7 \ \hbox{MeV} /r =0.85 \ \hbox{MeV}$.

Now we  estimate the time it takes
for the energy
deposited from the explosion of the pearls
to be lost by neutrino emission.
Crudely we take the {\em excess temperature $\Delta T$}
to dominate the whole temperature.
 At the temperature
of $\Delta T= 0.85 \ \hbox{MeV}= 8.5 * 10^{9} \ K$,
and using the density $10^{10} \ \hbox{g/cm$^3$}$
from \cite{Bethe,Cooperstein},
the rate of heat transport out of
the region by neutrinos becomes \ct{Woosley,Munakata}
$10^{24} \ \hbox{erg/cm$^3$/s}$, which
translates into $10^{14} \ \hbox{erg/g/s}$.
Now we must count that a major part of
the excess energy of $10 \ \hbox{MeV/nucleon}
= 10 \ \hbox{MeV}*6 *10^{23}/\hbox{g} =
10^{19}\ \hbox{erg/g}$
will have to be emitted in this way, but
that some of the energy
remains keeping the region above the ball somewhat
hotter (see points 4 and 5). Perhaps
part of the heat could even split the iron into nucleons
and/or helium (point 6). Also  some
of the heat falls into the neutron star, because
it does not manage to get out
before the genuine collapse of the star
takes place. Suppose we
take that half the energy (per gram),
i.e. $0.5 * 10^{19}\ \hbox{erg/g}$,
is properly emitted by the neutrinos.
Then with the neutrino energy loss rate of
$10^{14}\ \hbox{erg/g/s}$, this will take
$0.5*10^5\ \hbox{s} = 14\ \hbox{hours}$.
After this time relative to  the ignition of the
dark matter pearls,
the star will restart its gravitational collapse.

Let us assume that the first attempt by the star to
collapse after the silicon to iron burning era, which
got stopped by the pearl explosions, was accompanied
already by a significant neutrino burst
that, in the case of supernova SN1987A,
was observed by the LSD detector at Mont
Blanc \ct{MontBlanc}. In our picture, this
Mont Blanc neutrino burst was emitted
just before our pearls exploded.
Then we estimate, in a period of order 14 hours later, the
heated interior of the star would
cool down and the Kelvin-Helmholtz gravitational collapse
would restart after this delay. This next collapse
is supposed to be the main one responsible for
the neutrino burst seen by
Kamiokande and IMB \ct{Kamiokande1,Kamiokande2,IMB}
followed by the genuine
emission of the supernova remnants.
Both these neutrino bursts could have come from the very central
part of the star - meaning distances from the center
of the order of only a neutron star size of say 10 km.
The neutrino emission during the 14 hours period, on
the contrary,
would be so weak that there would be no
chance to see them experimentally on the
earth.

Both the above neutrino outbursts
involve large bunches of neutrinos,
because they arise from the
{\em deep} interior of the supernova.
We expect the energies or the temperature
of the neutrinos to be largest in the second of the two
outbursts. Let us argue for this expectation using a basically
oversimplified set of assumptions:
Counting only
the electron neutrinos,
which are produced from the protons by picking up
an electron from the degenerate gas of electrons
and becoming a neutron plus a neutrino.
Then, for material starting out with roughly similar
amounts of protons and neutrons which end up as
solely neutrons, the amount of neutrinos of this sort is
just proportional to the number of nucleons.
But now let us take the crude
approximation that
the radius of the resulting neutron star is
only weakly dependent on the amount of matter
in it. Then the potential energy released per nucleon in the
first collapse of say a mass $M_1$ would be
proportional to the average mass already fallen in
during the falling period, which means proportional
to $M_1/2$.
However the energy
released per nucleon during the fall in of
the next clump of matter of mass $M_2$ say would be
similarly proportional to $M_2/2 +M_1$.
Thus the temperature ratio for neutrinos in the
second collapse to that for those in the first
would be $\frac{M_2/2+M_1}{M_1/2} > 2 $.
For instance if $M_1 = M_2$,
this ratio would mean that the second
burst would deliver neutrinos with 3 times as
large a temperature.
The re-scattering or absorption and re-emission
of neutrinos, which ends with
them all coming from a certain
``optical depth'', may smooth away
part of this temperature difference.
In fact the fits \cite{deRujula, Kamiokande1,Kamiokande2} to
the Mont Blanc and Kamiokande
observations of the first and the second bursts
of neutrinos give temperature estimates
of 1 MeV and 4 MeV respectively.


The most remarkable coincidence
supporting
our whole model is that, in the case
of supernova SN1987A, {\em two bunches}
of neutrinos were indeed observed - one by the Mont Blanc
experiment
\cite{MontBlanc} and one by both
Kamiokande \ct{Kamiokande1,Kamiokande2}
and IMB \ct{IMB} - with a time difference of
4 hours 43 minutes.
Our 14 hours estimate for this time
difference
is off by a factor of 3. However the
potential difference between the
inside and outside of our dark matter pearls,
$\Delta V = 10 \pm 7\ \hbox{MeV}$, is
uncertain by a factor of 3 and our whole
estimate was anyway very
crude. So  we
could not hope for better agreement.

\subsection{Revival of the Shock
Wave}
\label{revival}
After the second collapse of the star the main effect
of our dark matter pearls will have disappeared.
{\em However a region mainly
a few hundred kilometers
away from the very center
has been heated up
significantly by the explosion of the pearls.} In this
region  the Fe peak
materials will even have split into nucleons or into
helium say.
%
%
The gravitational fall of the
core matter into the center causes a
compression to twice nuclear matter
density in a contracted proto-neutron
star object. The highly
contracted proto-neutron star rebounds and
very strongly expels the matter around
it. This causes the usually
expected shock wave to appear, below which the
matter moves
outward and above which the matter falls down.
Now the usual problem with models for supernova
explosions
is that detailed estimates of the
propagation of the shock wave
indicates that the shock wave stalls - at least in
``one-dimensional'', i.e. rotationally
symmetric, models. This means the shock wave does
{\em not} come out of the iron core and
fails to produce the energy needed to deliver an
outburst of supernova remnants.
This problem persists even when neutrino
transport of energy from the center to higher regions
in the iron core is included, as in standard simulations.
The explosive energy needed to
get sufficient supernova remnants for
matching with observations
is estimated to be about
$1 \ \hbox{foe} \equiv 10^{51} \
\hbox{ergs}$.
However, even the non-rotationally
invariant models with delayed neutrino heating
turn out not to
provide a full one foe for the supernova explosion.
Rather one typically only gets
a fraction, say 1/3, of a foe. So, in order to
realize a viable
model for supernovae, it seems to be required
that the three-dimensional or
two-dimensional
(meaning non-rotationally invariant)
treatment should somehow bring along an
extra boost reviving the shock wave.
In fact no current simulation using delayed neutrino heating
has produced a successful 1 foe supernova
explosion \ct{Papish}.

So, if it were not for our pearls, there
would again be
the same problem that the shock wave
would stall before
managing to provide the explosion of
the supernova.
Now the revival of the shock wave
by the energy available from the
recombination of elements,
say in the iron peak, which had earlier been
split into nucleons by the shock,
has recently been discussed in \ct{Papish,Fernandez}.
In the usual picture, this mechanism turns out
not to be so helpful for reviving the
shock wave and generating  an explosion with an energy of 1 foe.
The problem with the usual picture is that it needs
a flux of delayed neutrinos to bring the dissociated
or partially recombined nucleons (e.g.~to alpha
particles) up from $\lesssim 150\ \hbox{km}$ in height to
$ \gtrsim 500\ \hbox{km}$.
However our dark matter pearls can heat up the material
and/or split nuclei into nucleons all the way up to a height of 500 km.
So, in our picture, energy is deposited at the
height where it can help with
the explosion.
Also the explosions of our pearls can easily
provide energy for the revival of the shock wave.
For example the shock wave
would not have to split iron as it propagates, because
the iron would already have been split previously
by our dark matter explosion.

It is well-known that there is no supernova explosion
in a rotational invariant or 1D type calculation
\cite{Liebendoerfer} without a change in the physics.
However, with the {\em extra energy from our dark matter balls},
even the 1D approximation {\em can} generate an explosion.
Although, with our mechanism we do not need
convection to obtain an explosion, there are reasons
to believe that such convection, meaning a non-rotational
invariant explosion, is there anyway.

\section{Conclusion}
\label{sec5}
We have previously speculated that dark matter
consists of pearl-sized balls containing
a different type of vacuum - one with a condensate
of bound states of 6 top + 6 anti-top quarks -
and very strongly compressed ordinary matter.
We have here proposed that these dark matter
balls can become
active and suck in ordinary matter, if
they become surrounded by material with
a sufficient amount of free neutrons. The
activity of these pearl-sized
balls in a supernova
consists in first of all taking in
the free neutrons and thereby
expanding themselves to a bigger and bigger size.
Since the potential for nucleons in the
vacuum inside the pearls is supposed
to be 10 MeV lower for nucleons than outside, this expansion of the
pearls liberates 10 MeV energy for each
nucleon absorbed. The fast absorption of neutrons
makes the expansion explosive and produces a
large amount of energy in the region
up to, say, 500 km from the center.
This explosion
is supposed to stop or rather
postpone the usual
Kelvin-Helmholtz gravitational collapse
of the supernova, which begins at
the end of the era of silicon burning
to the iron peak elements.
Before it is halted the Kelvin-Helmholtz
collapse
already begins to
produce
a bunch of neutrinos which, in the
case of supernova SN1987A, was observed
as the ``first bunch'' of neutrinos
by the Mont Blanc experiment.

Then the interior of the star, heated by
the explosion of the dark matter pearls,
cools down by neutrino emission until
the gravitational collapse can restart
and generate a second bunch of neutrinos.
We estimated that this would happen a period
of order 14 hours after the interruption of
the first collapse.

Support for our model is provided by the fact
that, in the supernova SN1987A, there seemingly
were indeed {\em two bunches of strong neutrino
bursts} - each of a length of the order of 10 s.
Furthermore there was an interval of 4 hours 43
minutes between the two neutrino bursts, which
is perfectly consistent with our crude order of
magnitude estimate of 14 hours for this delay time.
A further important achievement of our model
is the provision of an extra source of energy
by the expansion of our dark matter pearls, which
is well suited to {\em  revive} the shock wave
expelled by a newly formed neutron star. This extra
energy is also able to deliver the observed 1 foe of energy needed
by the stellar remnants to escape.

The dark matter pearls start out from cm-size
with a density of order $10^{11}\ \hbox{g/cm$^3$}$.
However, in the presence of a supply of free neutrons,
the pearls rapidly expand until the (neutron) density in
the surrounding material becomes sufficiently low.
As the balls get larger the electric field surrounding the balls
gets weaker - although more extended - which allows
the balls more easily to glue together,
finally forming
one big ball surrounding the neutron star.

\section{Acknowledgment}
          One of us HBN wishes to thank
the Niels Bohr Institute for his
status as emeritus professor with an
office. CDF would like to
acknowledge
hospitality and support from Glasgow
University and the Niels Bohr Institute.

\end{document}